# Rethinking external validation for the target population: Capturing patient-level similarity with a generative model

Mohammad Azizmalayeri[1,2], Ameen Abu-Hanna[1,2], Saskia Houterman[3], Marije M. Vis[2,4], Giovanni Cinà[1,2,5], on behalf of the NHR THI registration committee

[1] Amsterdam UMC location University of Amsterdam, Dept of Medical Informatics, Amsterdam, The Netherlands

[2] Amsterdam Public Health Research Institute, Amsterdam, The Netherlands

[3] Netherlands Heart Registration (NHR), Utrecht, The Netherlands

[4] Amsterdam UMC, Dept of Cardiology, Amsterdam, The Netherlands

[5] Institute for Logic, Language and Computation, University of Amsterdam, Amsterdam, The Netherlands

## Abstract

**Background:** External validation is essential for assessing the transportability of predictive models. However, its interpretation is often confounded by differences between the external population and the development cohort, making it difficult to distinguish true model deficiencies from case-mix effects, as well as by limited consideration of the intended deployment environment. This study introduces a framework to address these limitations.

**Method:** We propose a framework that formalizes validation targets by distinguishing whether deployment is expected within the development or external population. The framework estimates these targets by quantifying each external patient's similarity to the development data and using these similarity scores to partition the external population into subgroups with varying levels of alignment to the development distribution. We use generative models, specifically autoencoders, to estimate similarity, offering a more flexible alternative to traditional linear approaches and enabling validation without sharing the original development data. The utility of autoencoder-based similarity measure is demonstrated using synthetic data, and the framework's application is illustrated using data from the Netherlands Heart Registration (NHR) to predict mortality after transcatheter aortic valve implantation.

**Results:** Our framework revealed substantial variation in model performance across similarity-defined subgroups, differences that remain hidden under conventional external validation yet can meaningfully alter conclusions. In several settings, conventional external validation suggested poor overall performance. However, after accounting for differences in patient characteristics, for some sub-groups, the model performance was consistent with internal validation results. Conversely, in some settings, apparently acceptable overall performance masked clinically relevant performance deficits in specific subgroups.

**Conclusion:** The proposed framework enhances the interpretability of external validation by linking model performance to population alignment with the development data. This provides a more principled basis for deciding whether a model is transportable, how it should be adapted, and to which patients it can be safely applied.

## 1. Introduction

Recent advances in machine learning (ML) models have sparked significant interest in developing prediction models for healthcare applications [1, 2]. However, a persistent challenge is ensuring that these models generalize effectively to new data encountered post-deployment [3, 4]. Model generalizability is typically assessed through internal and external validation studies [5, 6]. Internal validation evaluates performance using data from the same source as the development set, ensuring that the model performs well under new patients of the same population on which it was trained. Common approaches include splitting the dataset into development and validation subsets, or using resampling techniques such as cross-validation and bootstrapping [6].

External validation assesses the model on a new, different (but still relevant) dataset compared to the development set; for example, an external validation dataset could contain the same type of patients, say adult diabetic patients, but from a different hospital or geographic region. External validation serves two related purposes: when the model is intended for use within the development population, it assesses generalizability under potential distributional changes; when the model is applied to a different target population, it evaluates performance in that new population. Consequently, the interpretation of external validation depends on the intended target population—that is, the population in which the model is expected to be applied to in practice [7].

Interpretation of external validation results is further complicated by uncertainty regarding how, and to what extent, the external dataset differs from the development data. External data may closely resemble the development distribution or differ substantially, sharing the same variables but exhibiting different underlying population characteristics. This variability complicates interpretation of results because a single summary performance metric (such as the area under the receiver operating characteristic curve (AUC) or calibration slope) is often used to capture two different effects simultaneously: the model's true predictive ability in the population for which it was developed, and the impact of differences between the development and external populations [8]. As a result, when performance is worse in external validation than in internal validation, it becomes difficult to discern whether the model itself is flawed for its intended population, or if it is simply being applied to a population for which it was not designed.

Some approaches are suggested in the literature to address this issue. A common approach is quantifying the overall degree of relatedness between the development and validation datasets, which provides a summary measure of how similar the two datasets are [9]. However, this approach remains limited because the measure of performance is still influenced by the mixture of validation instances that either align with or deviate from the development data. In other words, the degree of relatedness does not disentangle the model's predictive performance on similar data from the effects of distributional changes. Recent studies have attempted to improve on this approach by employing propensity weighting to either align the distribution of the external dataset with the development distribution or to emphasize instances that are similar to the development cohort [10, 11]. However, this strategy has several limitations. First, it only provides insights into scenarios where the data resembles the development distribution, leaving performance under a different covariate distribution unexamined. Second, it does not account for the intended deployment target context: when the goal is to assess model performance within the external distribution itself, alignment with the development distribution may be less relevant. Finally, these methods typically rely on a linear probability-based similarity measure, which has been shown to be inadequate for capturing changes in data distribution [12], thereby limiting its effectiveness for robust alignment.

To mitigate ambiguity in interpreting external validation, we propose guiding it with nuanced estimands aligned with where the model will be deployed and which aspects of its behavior are most relevant [13, 14]. This approach enables defining key relevant validation questions and specifying the evaluation strategy needed to answer them. For example, when a model is intended for deployment within the original development population, poor performance on patients with data outside the development distribution, referred to as out-of-distribution (OOD) instances, may be less concerning, particularly if an OOD detection strategy is employed. In contrast, when the model is deployed in an external population, such patients constitute part of the new target population and are therefore directly relevant to assessing model performance [15].

To meet this need, we propose a more targeted and granular validation framework. Based on the questions that arise in the deployment context, we decompose the external dataset using patient-level similarity to the development data, stratifying it into subgroups such as "in-distribution-like (ID-like)" patients, whose data resemble the development set, and "OOD" patients, based on a quantitative measure of similarity or distance. This approach enables performance to be evaluated within subgroups that more directly address deployment-relevant questions**,** rather than treating the external sample as a homogenous whole. For instance, performance on the ID-like group reflects the model's intrinsic predictive validity, while performance on the OOD group reveals its robustness to covariate change. This approach turns external validation from a black-box summary into a more interpretable and diagnostic tool that moves beyond aggregate performance metrics.

In addition, we propose leveraging generative models, particularly autoencoders, to quantify patient-level similarity to the development data [16]. Unlike prior approaches that rely on probability-based similarity measures, this strategy provides more accurate distance estimates. More importantly, it enables validation without repeated access to the development data: an autoencoder can be trained once to learn the overall development data distribution and then used to measure similarity for multiple external datasets, whereas prior approaches required retraining a new similarity model for each external dataset.

Overall, this work makes three key contributions. First, we frame external validation as a deployment-driven process, emphasizing the need to distinguish intrinsic model performance from the effects of population differences (Section 2). Second, we introduce the use of generative models to quantify patient-level distance from the development data, improving upon prior linear probability-based approaches (Section 3). Finally, using data from the Netherlands Heart Registration (NHR), we conduct external validation experiments that illustrate how this framework improves understanding of model behavior under deployment (Section 4).

## 2. Framework for External Validation

This section presents a framework for validating prediction models using external data. We describe the key steps for conducting external validation under two deployment scenarios: (1) deployment within the development population and (2) deployment at the external population. For each scenario, we specify the key questions that external validation should aim to answer and formalize them as estimands. We then propose a data-driven stratification of the external dataset into subgroups to estimate these quantities and explain how the results can be contextually interpreted. Figure 1 provides an overview of the proposed framework, highlighting the methodological steps and interpretation strategies specific to each scenario.

We assume that the model is developed on data $(X, Y)$, where $X$ denotes the predictors and $Y$ the outcome, and validated on an external dataset $(X^*, Y^*)$. By definition, external data is assumed to be relevant to the development data; thus, we assume some overlap between the distributions of $X$ and $X^*$, although no underlying assumption is made regarding the extent of this overlap. Moreover, we keep our framework agnostic to the specific performance metric used; it can be applied with any appropriate measure that aligns with the clinical or analytical objective.

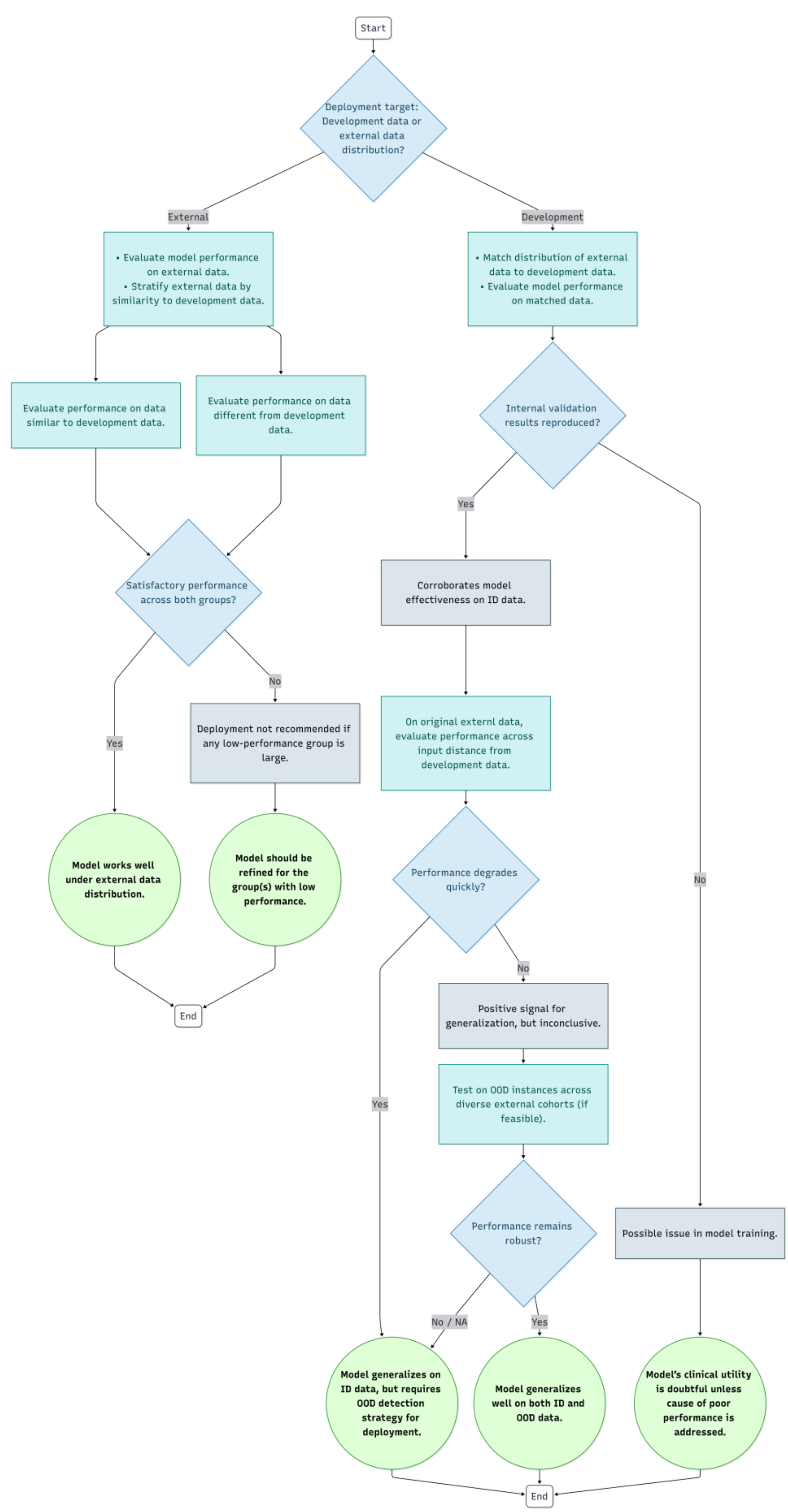

*Figure **1**. Flowchart illustrating the proposed external validation framework. Blue boxes indicate operations, gray boxes highlight key points, diamonds represent decision points, and circles denote final conclusions.*

### 2.1. Scenario 1: Deployment at the Development Population

In this scenario, the goal of external validation is to gain a deeper understanding of the model's predictive behavior for its intended use at the original development site. The external data is not the target population per se but serves as a probe to assess the model's robustness under slightly altered yet related conditions. The key questions that external validation should answer in this context are outlined below.

**Key Question 1: Does the model maintain similar performance on data matched to the development data distribution?**

Before evaluating performance on novel data, the model should first be tested for reproducibility—that is, to confirm it can replicate its results on data resembling the training distribution, which also represents the intended deployment target. This means removing the confounding impact of distributional differences from the analysis.

- **Methodology:** To assess performance under the development distribution, the external data distribution should be reweighted to match the development data. This can be achieved, for example, by applying propensity score weighting to the external instances so that their covariate distribution aligns with that of the development data [10, 17]. Details of the weighting procedure are provided in Appendix A. Model performance is then evaluated on the matched data and compared to the internal validation results.
- **Interpretation:**
  - Consistent Performance: Comparable results between matched external data and internal validation indicate that the model behaves consistently across data resembling the development distribution, supporting its reliability in such deployment settings.
  - Inconsistent Performance: Substantially worse results on matched external data may indicate potential model deficiencies. Since covariate differences are controlled for, such discrepancies could reflect limitations in the model, such as overfitting to the training data or concept drift (changes in $P(Y|X)$) [18, 19]. This outcome raises serious concerns about the model's reliability, even for its intended use at the development population. The model should undergo a more thorough diagnostic evaluation before any deployment is considered.

**Key Question 2: How robust is the model to OOD data?**

For safe and reliable deployment, it is crucial to understand how a model behaves when exposed to patients whose data distribution differs from that of the development cohort, an assessment also known as transportability.

- **Methodology:** We evaluate the model specifically on the subset of external data dissimilar to the development data. Ideally, this should not be treated as a binary check, but rather as an incremental analysis where performance is charted across varying levels of similarity to, or distance from, the development data using metrics such as the Mahalanobis distance [20]. The choice of an appropriate distance metric is discussed in Section 3.

- **Interpretation:**
  - Quick Performance Drop: A sharp drop in performance as data diverges from the development distribution indicates that the model is brittle and fails to generalize under covariate shift. In such cases, safe deployment requires identifying and flagging patients who differ from the development data, an approach known as OOD detection in ML [21, 22]. This is indeed under the assumption that the model itself remains fixed; improving its generalization would, of course, be a complementary solution.
  - Stable Performance: If performance remain robust and does not decline much as the data diverges from the development distribution, it suggests that the model has learned a more fundamental relationship between predictors and outcome that can generalize beyond the training setting. While encouraging, such results should be interpreted with caution: OOD populations are often heterogeneous and may not be fully represented in a single external dataset [23]. Robustness to covariate shift should ideally be verified across multiple diverse external cohorts before drawing strong conclusions about the model's generalizability.

**Overview of key questions:** Question 1 focuses on model performance when applied to data with the same distribution as the development cohort, whereas Question 2 examines how well the model generalizes when the data distribution differs.

### 2.2. Scenario 2: Deployment at the External Population

In this scenario, the external dataset represents the new target population, and the goal is to determine whether the model, as is, can be reliably transported and safely deployed in this new clinical environment.

The first step is the standard external validation, as the external dataset directly represents the target deployment environment. This provides the top-level answer regarding the model's potential utility for the new population. However, whether the performance is good or bad, a deeper diagnostic analysis is required to understand *why*, which is the focus of the next question.

**Key Question 3: How does performance vary across instances that are similar or dissimilar to the development data?**

To understand the model's behavior in the new environment, it is crucial to break down performance based on how closely each instance in the external dataset resembles the development data. This helps determine whether poor performance (if observed) stems from distributional differences between datasets or from intrinsic model limitations to reproduce the performance.

- **Methodology:** We stratify the external dataset into two conceptual groups: ID-like instances, which resemble the development data, and OOD instances, which are dissimilar. This stratification is derived from a similarity metric that quantifies each instance's distance from the development distribution, after which a threshold is applied to separate the ID-like and OOD groups. The choice of similarity metric is discussed in Section 3. Model performance is then evaluated within each group to reveal differences in predictive behavior.

  As in Question 2, performance can also be analyzed continuously across the similarity spectrum. However, because OOD cases are part of the actual deployment target in this scenario, explicitly isolating and evaluating them remains essential. Even when using a continuous approach, a threshold is ultimately needed to distinguish between similar or dissimilar instances. To streamline interpretation, we therefore recommend defining ID-like and OOD groups from the outset. A data-driven approach for selecting this threshold is described in Appendix B.
- **Interpretation:**
  - Satisfactory performance on both ID-like and OOD strata: This is the ideal outcome, indicating that the model is robust and transportable across both ID-like (similar) and OOD (dissimilar) groups, supporting safe deployment within the target setting. That said, it is important to note that this analysis does not offer insights into how the model might perform under further shifts relative to the external data distribution itself.
  - Poor performance on one or both of the ID-like and OOD strata: Poor performance on the OOD group suggests limited generalizability to patients who are part of the deployment population but differ from the training data. Poor performance on the ID-like group is more concerning, as it implies unreliable predictions even for cases similar to the training data; comparing it to performance on their nearest neighbors in the development data can offer further insights [24]. In any case, if the affected subgroup(s) are not negligible in size and pose a risk in deployment, the model should be refined, ranging from recalibration (if the issue lies in calibration) to full retraining to better represent those subpopulations.

**Overview of the key question:** Question 3 distinguishes model performance between instances that are similar to and those that are dissimilar from the development data within the external dataset. Unlike Question 1, the overall distribution of the external dataset is left unchanged. Although the methodology is similar to that used in Question 2, the interpretation is different.

## 3. Measuring Patient-Level (Dis)Similarity to the Development Data

Assessing the similarity of each external patient to the development data enables granular evaluation of model performance across ID-like and OOD regions. Existing linear discriminative approaches often fail to capture complex covariate shifts. We propose employing generative models as a more flexible and informative alternative.

**Existing approaches: Linear Discriminative Models**

A common approach to assess similarity between external and development datasets is to train a linear discriminative model, such as logistic regression, to distinguish between the two, often referred to as a membership model [9-11, 17]. The model's AUC quantifies overall separability, while predicted membership probabilities serve as instance-level similarity scores.

Although intuitive, this approach has key limitations. Its linearity constrains capturing real-world covariate shifts typically observed with complex and non-linear interactions. Moreover, discriminative models are trained to distinguish datasets rather than to model the structure of the development data itself. As a result, their similarity scores reflect features most useful for separation rather than true resemblance. Finally, membership models are dependent on the availability of the development data: each new external dataset requires a new membership model to be trained, thus requiring every time both development and external data.

**Alternative Approach: Generative Models**

Given the limitations of discriminative models, we advocate for a generative approach to measuring instance-level similarity. Generative models learn the underlying data structure, offering a more principled notion of measuring similarity of an instance to the data [25].

Among the generative models, the autoencoder has shown good capability in capturing data distribution for medical tabular datasets [12, 26, 27]. An autoencoder is an unsupervised neural network trained to compress data into a low-dimensional latent space and then reconstruct it [28]. When presented with unfamiliar instances, reconstruction quality deteriorates. This property can be exploited to define patient-level similarity: by training an autoencoder on the development dataset and measuring the reconstruction error for a new patient data, we can obtain a quantitative proxy for how (dis)similar that patient is with respect to the development population.

Autoencoders can capture non-linear relationships among features and are trained in an unsupervised manner using only the development data. As a result, they do not require labeled external data for training. A single trained autoencoder can therefore be reused to evaluate multiple external dataset, whereas discriminative approaches must be retrained for each new external dataset because they rely on learning a boundary between the development and external samples. It is important to train the autoencoder on a subset of the development data that is separate from the portion used to train the downstream predictive model to avoid introducing bias in the evaluations; for instance, the AE could be trained on the internal test set of a train/test split.

**Supporting Evidence: Simulation Study**

To empirically compare these approaches for detecting covariate shift, we conducted a simulation study using logistic regression and an autoencoder. A simple 2-dimensional dataset was drawn from a multivariate normal distribution to serve as the development data. Each model was then used to assign a similarity score  to points across the entire input space. Logistic regression required a contrastive dataset for training, so we generated a second dataset with shifted means and covariances to serve as the contrasting (external) data. The autoencoder, in contrast, was trained solely on the development data, reflecting its unsupervised nature. This setup enabled direct visualization of how each method defines similarity and responds to distributional shifts.

Figure 2 illustrates the results. The top row illustrates the sampled datasets for each case. The bottom row displays the similarity scores across the input space visualized using a color gradient (lighter = more similar, darker = less similar). The autoencoder effectively captures the nonlinear structure of the development data, with similarity scores degrading smoothly away from the data manifold, yielding an intuitive notion of dissimilarity. Logistic regression, however, imposes linear boundaries between similar and dissimilar regions which do not reflect true data geometry, sometimes assigning higher similarity to points farther from the distribution. This highlights its limitations in capturing complex or nonlinear covariate shifts.

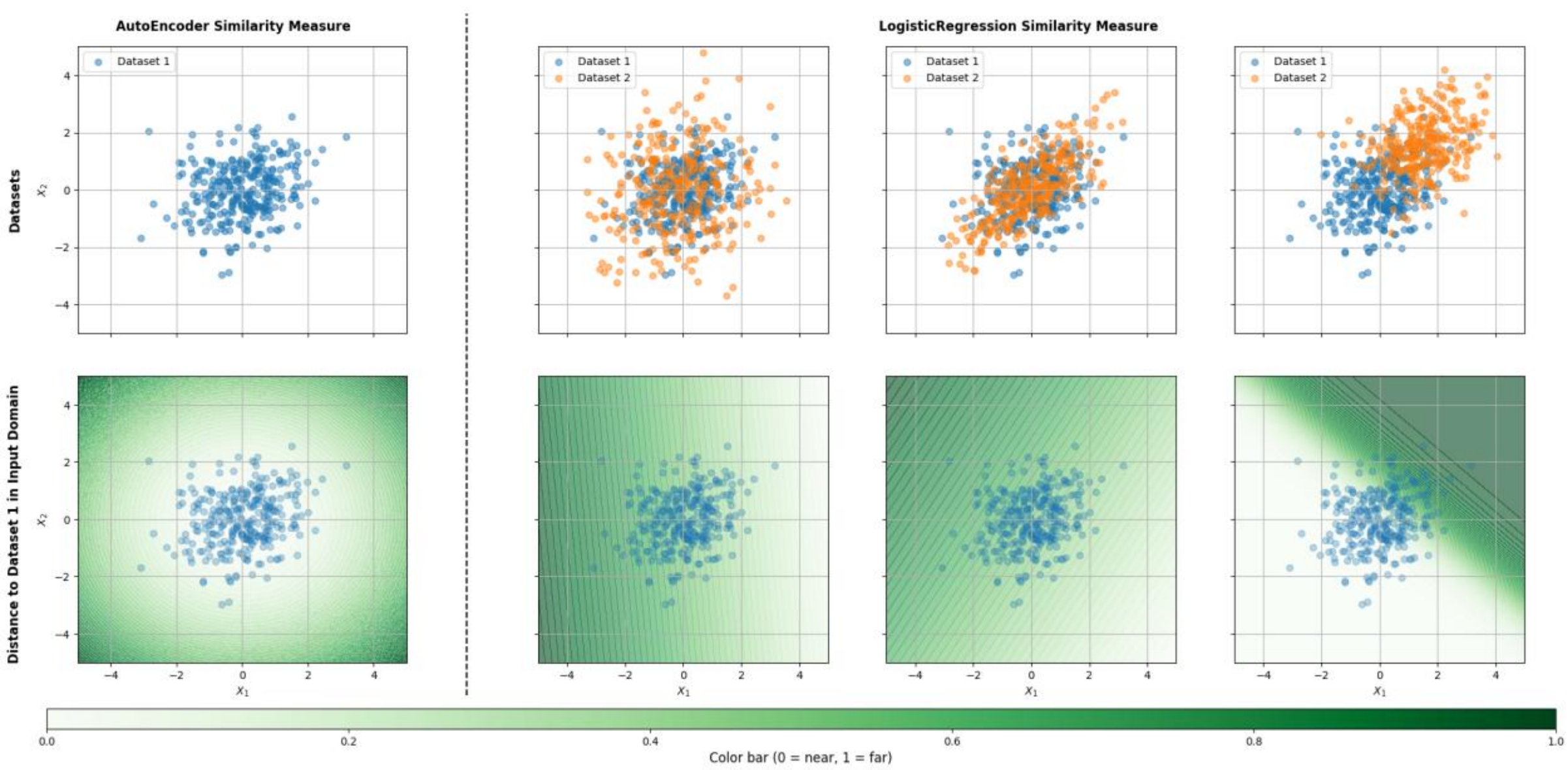

*Figure 2. Comparison of similarity scoring using an autoencoder versus a logistic regression. The top row shows the sampled datasets: Dataset 1 (blue) represents the development data to which similarity is measured, while Dataset 2 (orange) serves only as a negative class for training the logistic regression models. The bottom row displays instance-level similarity scores across the 2D input space using a color gradient (lighter = higher similarity). The autoencoder (first column) captures the nonlinear structure of the development data, with similarity degrading smoothly away from data manifold. In contrast, logistic regression (remaining columns) results in unintuitive scoring in regions far from the data.*

## 4. Case Study: Applying the Framework to Mortality Prediction

To demonstrate our proposed validation framework, we apply it to a case study involving 30-day mortality prediction following a transcatheter aortic valve implantation (TAVI), a minimally invasive procedure for patients with severe symptomatic aortic stenosis who are either ineligible for traditional surgical aortic valve replacement or are considered high-risk surgical candidates. This clinically relevant use case illustrates how our approach can yield deeper insights into model performance.

For this case study, we used the TAVI dataset collected by the Netherlands Heart Registration (NHR) [29]. The NHR is a nationwide, physician-driven and patient-focused quality registry that contains procedural and outcome data of all invasive cardiac interventional, electrophysiological and cardiothoracic surgical procedures from all Dutch hospitals. Data collection and registration is performed by the participating centers from the electronic health records in a secured online environment according to a detailed data dictionary. The dataset includes 23,418 patients across 16 centers admitted between January 1st, 2013, and December 31st, 2023. The multicenter structure of the data enables external validation by supporting model development on a subset of hospitals and evaluation on the remainder. Further details on data acquisition and analysis by the NHR can be found in recent publications [29-31].

The study was approved by the institutional review board MEC-U (W19.270) and conducted in agreement with the principles of the Declaration of Helsinki. A waiver for informed consent for analysis with the data of the NHR data registry was obtained.

**Setup**

We selected the 5 hospitals with the highest number of patients from the dataset and applied a leave-one-center-out cross-validation strategy to define the development and validation populations. In each iteration, one hospital served as the external validation site, while the remaining 4 formed the development set.

Within each iteration, data from the 4 development hospitals were randomly split into training (80%) and test (20%) subsets. The training set is used to develop a logistic regression model for 30-day mortality prediction, and the test set is used for internal validation. The developed model was then externally validated on the held-out hospital.

Pre-processing steps and variables selection followed a recent study on the generalizability of TAVI prediction models [32]. For example, Multiple Imputation by Chained Equations (MICE) was used to impute missing values, with the imputation model fitted only on the training data from the development hospitals to prevent data leakage. Table 1 lists the selected variables along with their average values across centers.

Using this setup, we conducted validation for two deployment scenarios: (1) deployment within the development hospitals and (2) deployment within the external validation hospital. For each case, we addressed the relevant questions posed by our validation framework and interpreted the results accordingly, illustrating how this approach yields a more nuanced understanding of model performance and potentially also different implementation decisions.

We employ an autoencoder-based similarity measure to quantify the distance of external instances from the development data. Moreover, the Brier score is used as the primary performance metric due to its additivity property over disjoint subsets of data. Specifically, as shown in Appendix C, if a dataset $D$ is partitioned into $n$ disjoint subsets $D_1, D_2, \ldots, D_n$, the overall Brier Score can be expressed as a weighted average of the Brier scores over subsets: $Brier(D) = \sum_{j=1}^{n} \alpha_j \, . Brier(D_j)$, where $\alpha_j = \frac{|D_j|}{|D|}$ is the proportion of instances in subset $D_j$. This property facilitates interpretation of model performance across stratified subsets. The AUC does not have this property, but we include AUC results for ID-like and OOD groups in Appendix D for completeness.

**Results: Deployment at the Development Hospitals**

Deployment at the development hospitals raises Question 1 (reliability under development distribution) and Question 2 (reliability under data shift). To address Question 1, we evaluate the model's performance on the matched external data with the development data in Figure 3. For context, we also include conventional vanilla internal and external validation results. To address Question 2, we plot model performance across deciles of similarity scores relative to the development data in Figure 4.

Figure 3 reveals that performance measured on matched external data yields more nuanced and different insights than conventional external validation. In settings where Hospitals 1, 3, 4, or 5 serve as the external validation site, standard validation approaches show a gap between internal and external validation performance, which is often interpreted as a negative sign for model performance. However, this gap diminishes or disappears for Hospitals 1, 3, and 4 when evaluated on matched data. This suggests that the observed gap in standard validation for these cases is largely due to distributional differences, and that the model remains reliable when applied to data similar to the development set. In contrast, the gap remains more pronounced for Hospital 5, indicating that the model may be intrinsically less reliable than what was observed in internal validation. Possible explanations for this behavior are discussed in Appendix E.

Figure 4 illustrates how performance varies as data diverges from the development distribution. The increase in prediction error with higher dissimilarity scores is observed across all settings; however, it is more pronounced in settings where Hospitals 1, 3, or 4 serve as the external validation site. For example, in the case of Hospital 4, prediction error increases gradually with data deviation but shows a sharp rise in the final decile. This suggests that the model's performance remains reasonable up to that point, but deteriorates significantly beyond it.

These insights can meaningfully alter the conclusions drawn from external validation. For instance, consider the model externally validated on data from Hospital 4 (developed on data from Hospitals 1, 2, 3, and 5). Conventional external validation suggests poor transportability, implying the model may be unsuitable for deployment. However, our analysis reveals that this performance gap largely reflects distributional differences rather than model failure. The model remains suitable for deployment within the development hospitals, provided it is accompanied by an OOD detection mechanism to flag and withhold predictions for highly dissimilar cases that may occasionally arise.

*Table 1. Characteristics of patients in TAVI dataset for the centers included in the analysis. Abbreviations: TAVI, transcatheter aortic valve implantation; LVEF, left ventricular ejection fraction; sPAP, systolic pulmonary artery pressure; NYHA, New York Heart Association classification; BMI, body mass index; BSA, body surface area; sCreat, serum creatinine; eGFR, estimated glomerular filtration rate.*

| | **5 centers of TAVI dataset** | | | | | |
|---|---|---|---|---|---|---|
| **Center** | **All 5** | **1** | **2** | **3** | **4** | **5** |
| Number of patients | 10693 | 2672 | 2203 | 2141 | 2005 | 1672 |
| 30-day-mortality (%) | 3.3 | 2.8 | 3.2 | 3.8 | 3.9 | 2.7 |
| Age years (mean) | 79.3 | 79.7 | 79.7 | 78.5 | 80.0 | 78.6 |
| Gender Male (%) | 52.8 | 50.8 | 54.3 | 53.8 | 52.1 | 53.3 |
| LVEF (mean) | 50.6 | 48.8 | 50.1 | 52.5 | 50.1 | 52.6 |
| sPAP mmHg (mean) | 29.9 | 29.2 | 29.3 | 35.4 | 29.7 | 28.8 |
| NYHA class 3 (%) | 48.8 | 51.5 | 44.3 | 46.7 | 64.6 | 33.5 |
| BMI kg/m2 (mean) | 27.1 | 26.9 | 27.0 | 27.6 | 26.5 | 27.7 |
| BSA threshold (%) | 56.0 | 56.8 | 53.7 | 56.2 | 57.0 | 56.2 |
| sCreat mmol/L (mean) | 105.7 | 105.6 | 108.0 | 107.1 | 107.1 | 99.4 |
| eGFR mL/min/1.73m2 (mean) | 61.0 | 61.7 | 59.0 | 61.1 | 60.0 | 63.8 |
| No Diabetes (%) | 73.1 | 77.1 | 73.1 | 68.8 | 74.2 | 70.6 |
| Previous cardiac surgery (%) | 18.1 | 13.8 | 23.9 | 15.6 | 21.2 | 16.3 |
| Chronic lung disease (%) | 18.7 | 21.2 | 16.3 | 17.7 | 19.1 | 18.4 |
| Critical preoperative state (%) | 0.3 | 0.3 | 0.2 | 0.2 | 0.2 | 0.5 |
| Recent myocardial infarction (%) | 1.8 | 2.2 | 2.2 | 1.2 | 1.1 | 2.6 |
| Access route transfemoraal (%) | 83.9 | 80.2 | 95.3 | 93.8 | 88.4 | 55.8 |

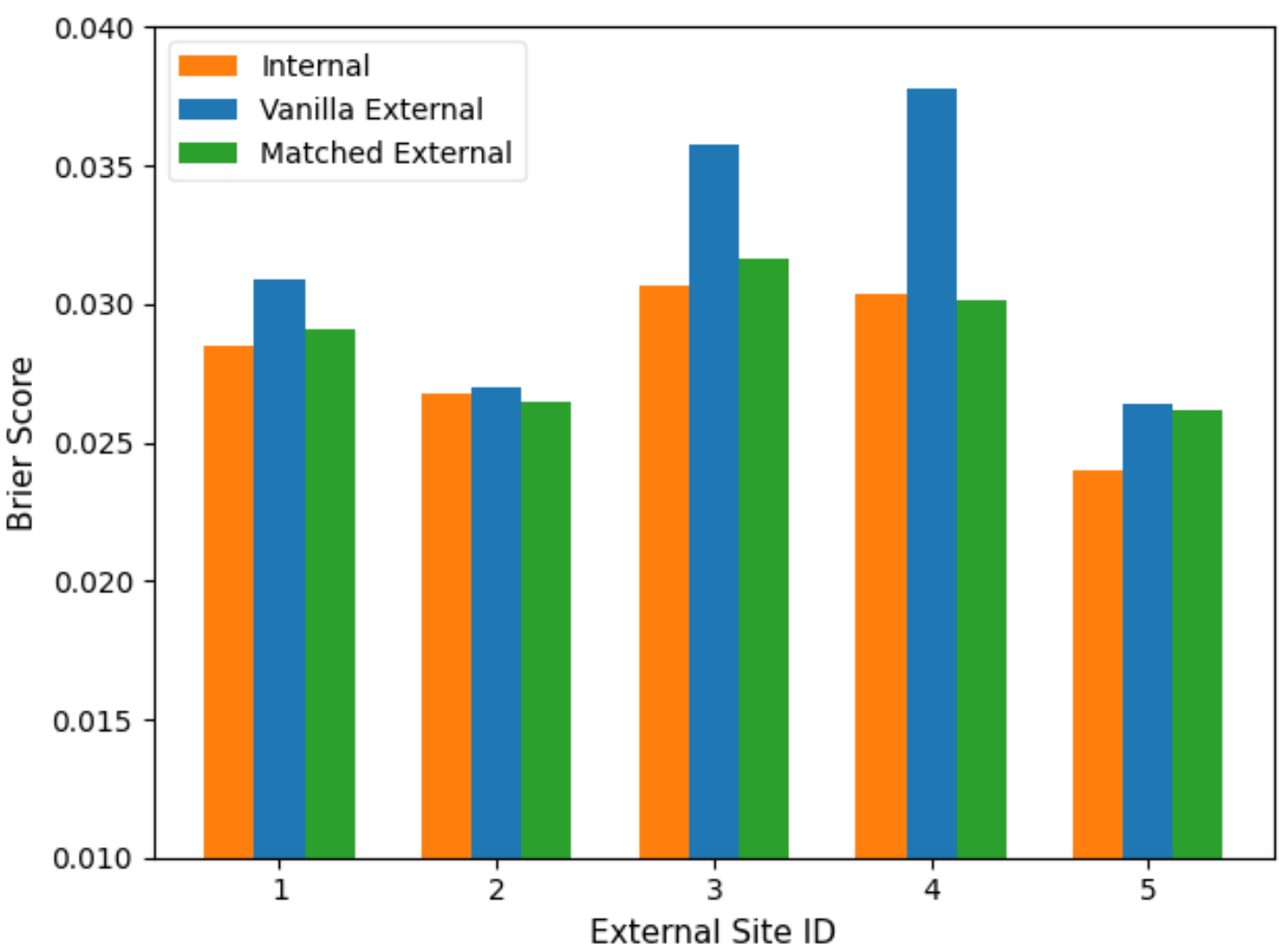

*Figure 3. Model performance on external instances matched to the development distribution (green bars), addressing Question 1. Standard internal and external validation results are included for comparison (orange and blue bars). Performance is measured using the Brier score, where lower values indicate better performance. The hospital used as the external validation site in each cross-validation fold is indicated on the x-axis.*

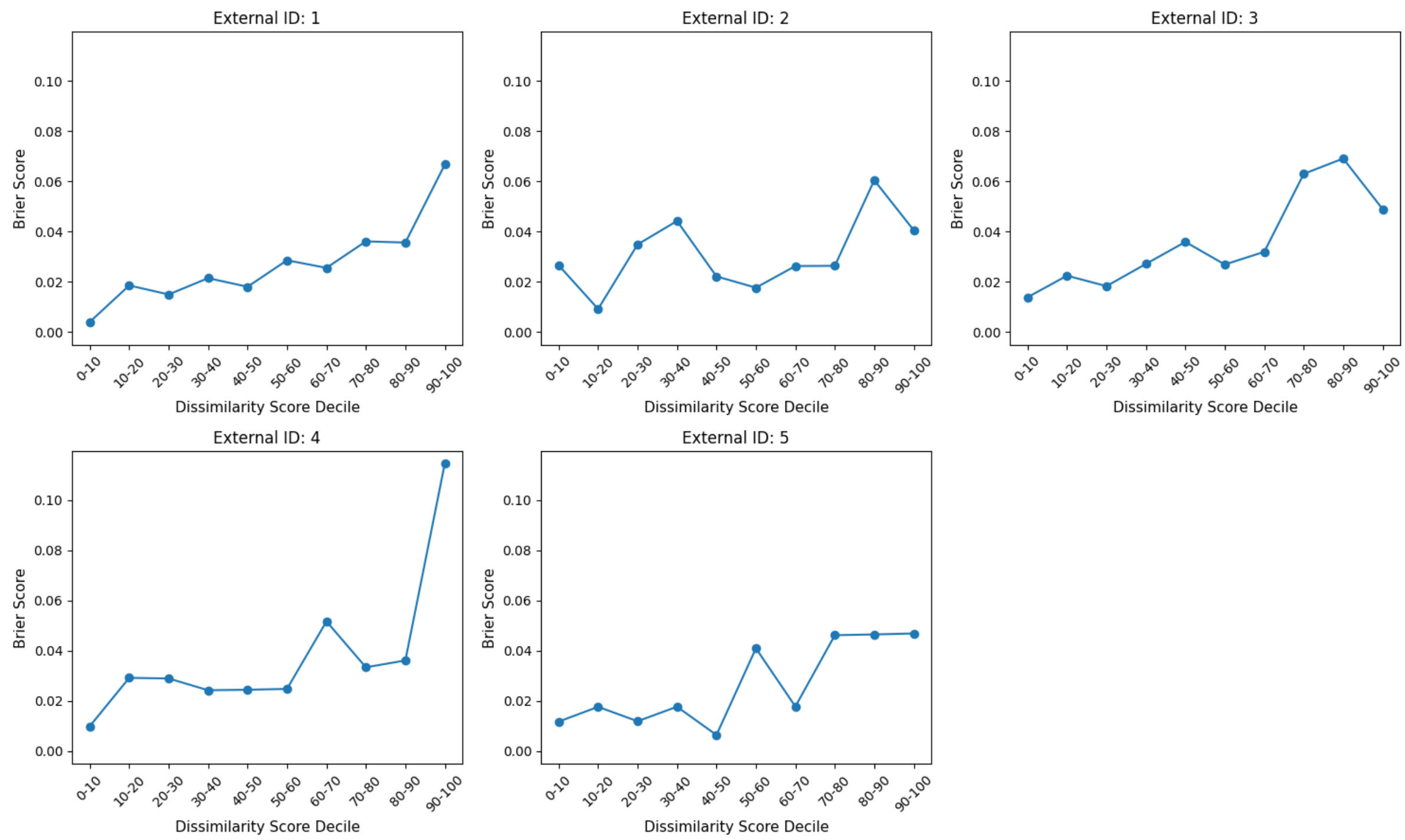

*Figure 4. Model performance on external data stratified into deciles (ten equally sized subgroups) based on similarity to the development data, addressing Question 2. The 0–10 decile represents the most similar cases, while the 90–100 decile includes the most dissimilar. This illustrates how performance varies as data diverges from the development distribution. Performance is measured using the Brier score, and the hospital used as the external validation site in each cross-validation fold is indicated in the plot title.*

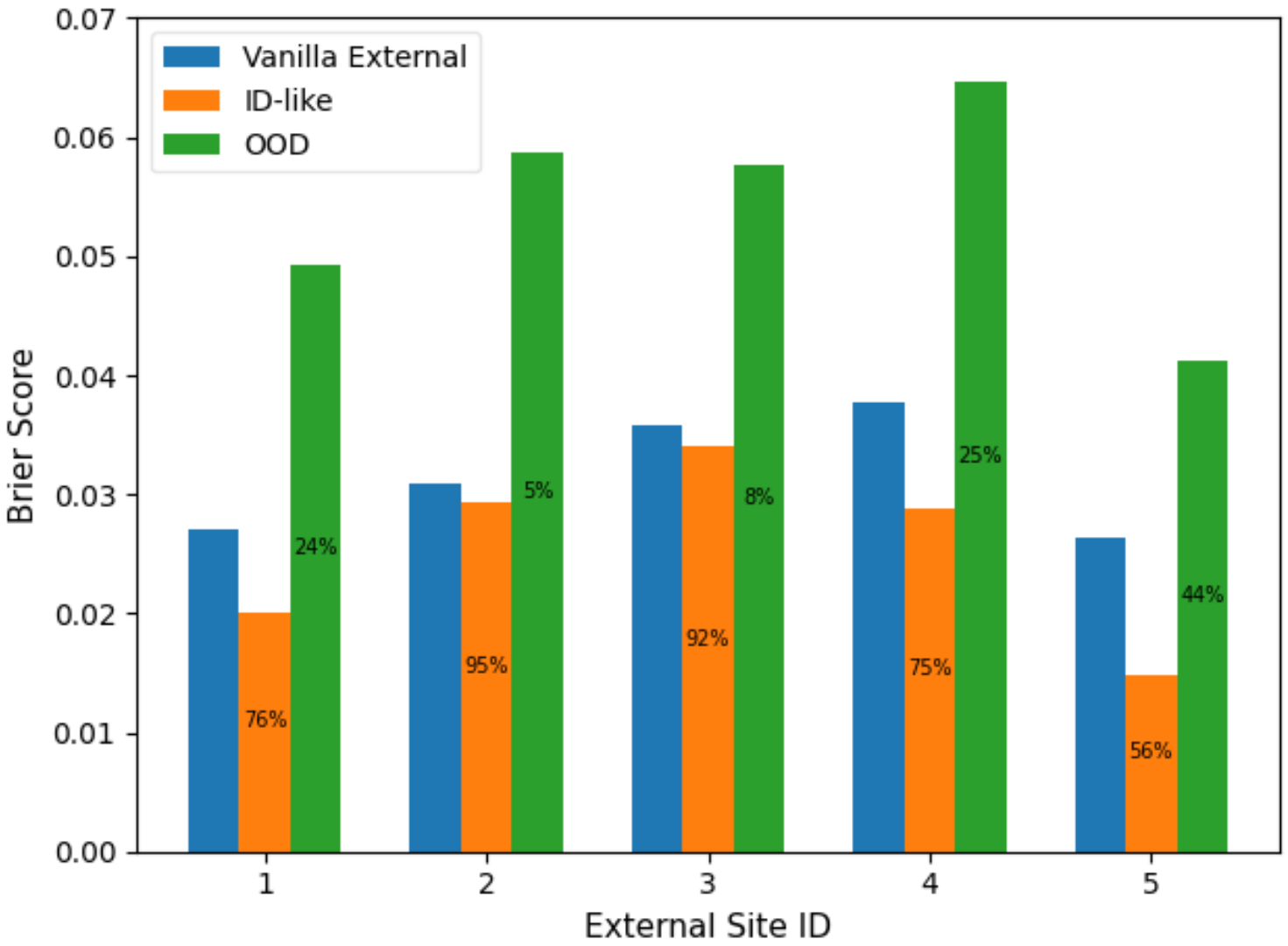

*Figure 5. Model performance on ID-like and OOD subgroups within the external validation site, addressing Question 3. The performance on external data (blue bars) is split into instances similar to the development data (ID-like, orange bars) and dissimilar instances (OOD, green bars), with the proportion of each subgroup represented on the bar. Performance is measured using the Brier score, where lower values indicate better performance. The hospital used as the external validation site in each cross-validation fold is indicated on the x-axis.*

**Results: Deployment at the External Hospital**

Deployment at the external hospital raises Question 3 regarding model performance on instances that are similar (ID-like) or dissimilar (OOD) to the development data. To address this, we report results for both subgroups, defined using the 90th percentile of the similarity score, in Figure 5. Conventional vanilla external validation results are also included for comparison, representing the weighted average of performance across the ID-like and OOD subgroups.

Figure 5 highlights that conventional external validation masks two substantially different performance values. In all cases, performance in the OOD subgroup is markedly worse than in the ID-like subgroup. This distinction changes how validation results should be interpreted and what corrective actions are warranted, depending on the magnitude of the performance gap and the prevalence of each subgroup in the deployment population. For example, in Hospitals 1, 4, and 5, a substantial proportion of patients fall into the OOD group, where error rates are considerably higher. Because these patients form part of the deployment target, the model should ideally be adapted to them to avoid delivering systematically poorer predictions for a large patient segment. In contrast, in Hospitals 2 and 3, the OOD subgroup is relatively small, allowing the model to be deployed as is, provided an OOD detection mechanism is in place to flag such cases and prevent unreliable predictions. These insights reveal opportunities for model recalibration and subgroup-specific adaptation, or alternatively for safe deployment with OOD filtering, dimensions that conventional external validation alone would not have captured.

## 5. Discussion

In this study, we introduced a framework to enhance the interpretability of external validation for clinical prediction models by explicitly accounting for the intended deployment target and the similarity between external data instances and the original development distribution. By quantifying instance-level similarity with an autoencoder, the framework decomposes the external dataset into subgroups based on their proximity to the development data, revealing how model performance changes as data diverge from the original distribution.

Applied to models predicting mortality after TAVI, this approach uncovered important variation in performance that conventional external validation would have obscured, leading to different conclusions on the model's viability. For instance, this approach revealed that performance under distributional change was markedly worse than suggested by standard validation in some cases, exposing risks that would otherwise remain undetected. This directly calls for additional actions, such as model adaptation or OOD filtering, when deploying in environments that differ from the development setting, or conversely indicate that, although conventional external validation may suggest limited deployability, the model can still be used within the development environment where distributional shifts are unlikely to arise frequently.

Our work aligns with and extends previous literature emphasizing that external validation results can be confounded by differences in patient characteristics between development and validation cohorts [7-11]. Earlier approaches, such as quantifying the degree of relatedness between development and validation populations or aligning the external dataset's distribution with that of the development set, address this challenge only partially [9-11]. They overlook the intended deployment population, rely on population-level adjustments without accounting for model robustness under distributional changes, and remain limited in capturing patient-level similarity to the development data because they rely on linear discriminative approaches. Moreover, they require access to the original development data for external validation. In contrast, our framework explicitly incorporates the intended deployment distribution, leverages a generative model to quantify instance-level similarity, evaluates performance across subgroups similar or dissimilar to the development data, and enables validation using only the trained autoencoder rather than the original data. The last point particularly means that the generative model can be trained only once and shared for validation, as opposed to the aforementioned linear methods that require repeated access to the development data, thereby mitigating privacy concerns about sharing the development data. In this way, it complements and advances the literature on evaluating the generalizability of clinical prediction models.

Despite these contributions, several limitations should be acknowledged. First, assessing performance under data changes or through matched datasets may not always be meaningful. For instance, evaluation under distributional change is irrelevant if the external data closely matches or is even a subset of the development data, while matching is not informative if there are insufficient instances in the external data that represent the development distribution. Assessing population-level differences can serve as a useful pre-check for the relevance of these analyses. Second, the absence of a clear cutoff for distinguishing instances “similar” or “different” from the development data makes it difficult to define a precise boundary between them. We addressed this challenge by offering a data-driven approach to define a threshold and by reporting performance across the continuum of similarity scores. Third, our results are based on an autoencoder, which admittedly is a simple generative model; whether these results generalize to more capable generative models is left to future work. Even so, one may argue that obtaining such results with a simple generative model is a testament to the fact that the approach has intrinsic potential. Finally, although generative models offer a powerful way to capture complex patient-level representations, their performance is contingent on the quality and representativeness of the data used for training. In practice, this may limit their applicability in settings with small sample sizes or highly heterogeneous populations.

For clinicians and decision-makers, this framework provides practical guidance on how to interpret external validation results in the context of deployment. Rather than relying solely on an aggregate performance estimate, they can assess whether observed performance variations stem from variations in patient populations or from limitations inherent to the model. This distinction is critical for clinical deployment: if performance decline is primarily due to differences in patient populations, strategies such as out-of-distribution detection, model adaptation, or recalibration may suffice. Conversely, if performance is inconsistent even under data similar to the development distribution, this points to more fundamental limitations of the model.

Future research should refine this framework and test it in other clinical domains and datasets to further assess its usefulness and robustness. In addition, when future data is conceived as ‘external’ data, the framework could be leveraged to guide strategies for model updating and adaptation, providing a structured way to decide when and how to recalibrate or retrain models according to the measured predictive performances across subgroups. Incorporating prospective validation studies will also be essential to evaluate how the framework performs in real-world deployment settings. Finally, while our approach focuses primarily on predictive performance, important future work lies in integrating it with broader considerations such as interpretability, fairness, and clinical utility, which are critical for the safe and effective adoption of prediction models in practice.

## Acknowledgments

The authors would like to thank Sagar Simha, Otto Nyberg, and Juliette Ortholand for their valuable feedback and suggestions on this manuscript.

**Funding**: This research did not receive any specific grant from funding agencies in the public, commercial, or not-for-profit sectors.

# Appendix

## Appendix A. Propensity Score Weighting for Distribution Matching

When evaluating model performance on external datasets, differences in covariate distributions between the development data and the external data may introduce bias. To address this, propensity score weighting can be used to reweight the external dataset so that its covariate distribution aligns with that of the development dataset.

Let $D_{dev}$ denote the development dataset and $D_{ext}$ the external dataset. Additionally, assume we have a similarity metric $f_D(x_i)$ that quantifies similarity of instance $(x_i, y_i)$ to dataset $D$. As described in Section 3, we recommend using generative models to estimate this similarity metric, rather than discriminative approaches such as logistic regression.

To reweight the external dataset so that its distribution matches the development dataset, each external instance is assigned a weight

$$\omega(x_i) = \frac{f_{D_{dev}}(x_i)}{f_{D_{ext}}(x_i)}, \qquad \text{for } (x_i, y_i) \in D_{ext}.$$

For instances in the development dataset, the weight is set to 1:

$$\omega(x_i) = 1, \qquad \text{for } (x_i, y_i) \in D_{dev}.$$

Any performance metric (e.g., Brier Score, calibration error, AUROC) can then be computed on the weighted external dataset. Many standard machine learning libraries, such as scikit-learn, provide support for instance weights when evaluating performance metrics.

Limitations. While propensity score weighting reduces distributional differences, it does not guarantee perfect matching. A perfect alignment requires that the external dataset includes sufficient coverage of the covariate space observed in the development dataset. If certain regions of the development distribution are absent (or underrepresented) in the external data, weighting cannot recover them, and performance estimates may remain biased. Moreover, the validity of the approach depends on how well the similarity metric is estimated.

## Appendix B. Threshold for Distinguishing Between Similar and Dissimilar Instances

A key challenge in separating similar from dissimilar instances lies in defining a principled threshold on the measured similarity scores, which depends on how strict one wishes to be in labeling an instance as similar. This challenge is further complicated by the fact that biomedical data often exhibit substantial heterogeneity across populations and conditions [33, 34]. Consequently, the distinction between “similar” and “dissimilar” instances is rarely binary and is inherently dependent on the context and level of variation considered relevant for the application.

In this work, to determine this threshold in a data-driven manner, we examine the distribution of similarity scores within the development dataset itself. Specifically, we compute similarity scores for each development instance (compared to the entire development set) and define a threshold based on a chosen percentile—for example, the 90th percentile of these scores. This ensures that any external instance labelled as "similar" is at least as close to the development data as a selected proportion of the original training instances. The chosen percentile can be adjusted according to the desired level of strictness in defining similarity.

**Appendix C. Decomposition of the Brier Score Over Disjoint Subsets**

Let $D$ denote a dataset of size $|D|$, and suppose it is partitioned into $n$ disjoint subsets $D_1, D_2, \ldots, D_n$. The Brier Score for dataset $D$ is defined as

$$Brier(D) = \frac{1}{|D|} \sum_{(x_i, y_i) \in D} (p_i - y_i)^2,$$

Where $y_i \in \{0, 1\}$ is the true outcome and $p_i$ is the predicted probability of the positive class for instance $x_i$. Since the subsets are disjoint and partition $D$, we can separate the sum and rewrite it as:

$$Brier(D) = \frac{1}{|D|} \sum_{j=1}^{n} \sum_{(x_i, y_i) \in D_j} (p_i - y_i)^2 = \sum_{j=1}^{n} \frac{|D_j|}{|D|} . \frac{1}{|D_j|} \sum_{(x_i, y_i) \in D_j} (p_i - y_i)^2 \ .$$

The inner term is exactly the Brier Score for subset $D_j$. Therefore, we can rewrite it as:

$$Brier(D) = \sum_{j=1}^{n} \alpha_j . Brier(D_j), \quad where \quad \alpha_j = \frac{|D_j|}{|D|}.$$

This shows that the Brier Score over the entire dataset can be expressed as a weighted average of the Brier Scores computed over disjoint subsets, with weights equal to the relative subset sizes.

**Appendix D. AUC Results**

In the main text, we reported results using the Brier Score loss, chosen for its additivity property as explained in Appendix C. For completeness, Figure S1 presents the corresponding results of Figure 5 when evaluated with AUC instead of the Brier Score. Since AUC does not possess the additivity property, only the ID-like and OOD results are shown. This figure further highlights the performance differences observed between the ID-like and OOD groups.

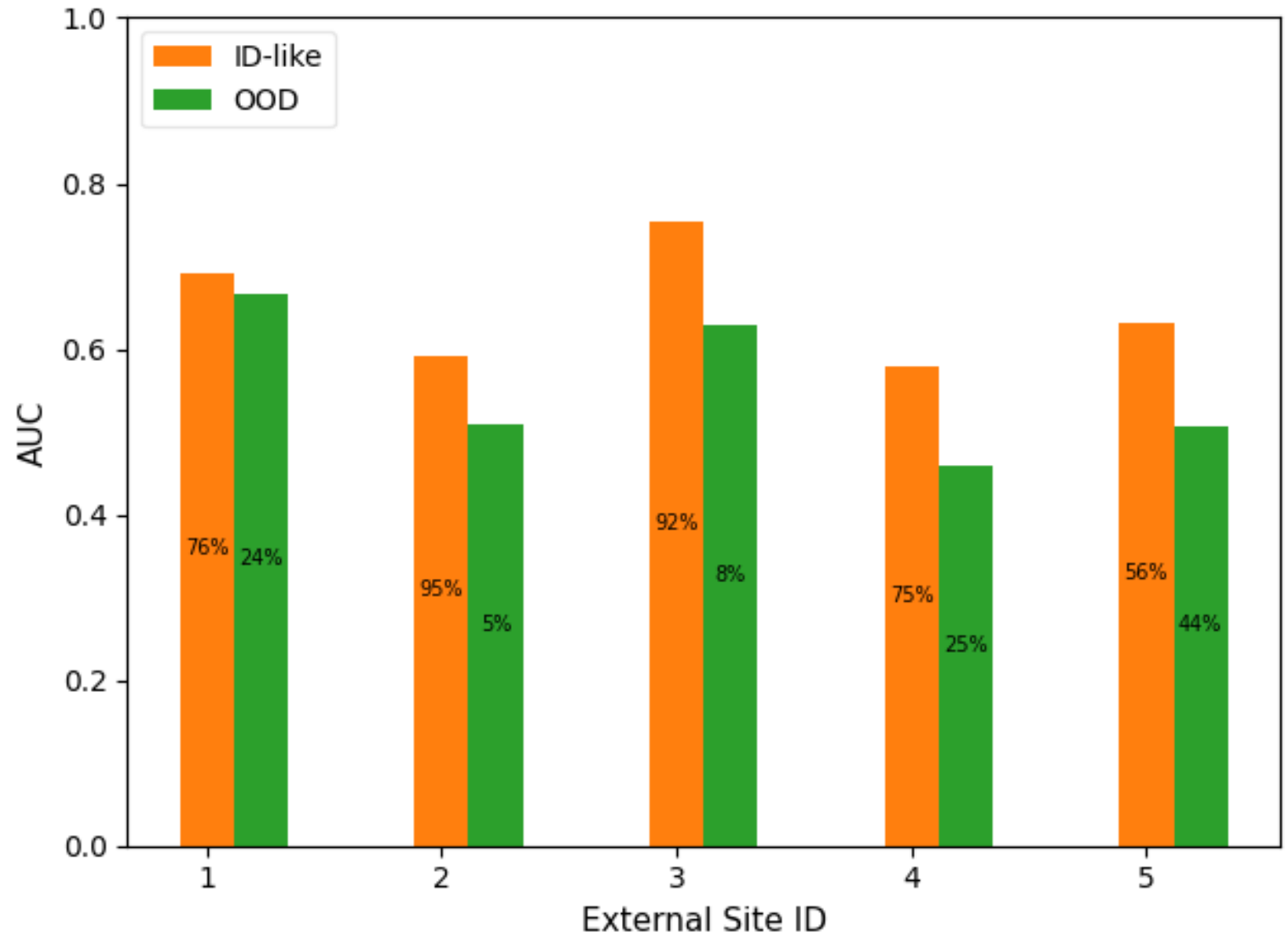

*Figure S1. Model performance on ID-like and OOD subgroups within the external validation site. Performance is measured using the AUC, where higher values indicate better performance. The hospital used as the external validation site in each cross-validation fold is indicated on the x-axis, and the proportion of external data in each subgroup is shown on the corresponding bar.*

## Appendix E. Explaining the Gap Between Internal and Matched External Validation

In the experiments described in Section 4, particularly in Figure 3, we observed that when Hospitals 1–4 were used as development centers and Hospital 5 served for validation, a gap emerged between internal validation performance and (matched) external validation performance. As shown in Table 1, Hospital 5 has a lower mortality rate (2.7%) compared to the other hospitals (3.4%). Even after applying weighting, the adjusted mortality rate in Hospital 5 is 2.8%, still noticeably different from the 3.4% in the development set. This indicates that matching the external data based on predictors does not fully align the distribution of outcomes. Consequently, the conditional probability $p(y|x)$ may differ in Hospital 5 relative to the other hospitals, potentially explaining the observed performance gap. In such a scenario, the model would remain reliable for the population on which it was developed but would not be robust to changes in $p(y|x)$.

To clarify further, as discussed in Appendix A, the distribution-matching weights are calculated solely based on the predictors $x$, without incorporating the true outcome $y$. This means that the matching procedure ensures similarity in the distribution of the predictors across datasets, but it does not guarantee alignment in the conditional relationship between predictors and outcomes, i.e., $p(y|x)$. As a result, even with perfect predictor matching, the outcome distribution in the external dataset may still differ from that of the development set. This discrepancy provides a plausible explanation for why the model's performance in internal validation does not fully carry over to the matched external validation. In other words, while weighting can correct for covariate shifts (differences in $p(x)$), it cannot correct for differences in the underlying outcome mechanism $p(y|x)$, which ultimately affects model generalizability.

Ideally, for a more rigorous assessment of model generalizability, one would want not only predictor matching but also outcome matching—ensuring that both $p(x)$ and $p(y|x)$ are aligned across datasets. However, since outcome alignment cannot be enforced through predictor-based weighting, differences in $p(y|x)$ remain a source of potential performance gaps.

### Appendix F. Members of the NHR THI Registration Committee

Table S1 lists the members of the NHR Transcatheter Heart Valve Interventions (THI) Registration Committee.

*Table S1. Members of the Transcatheter Heart Valve Interventions committee of the Netherlands Heart Registration Committee.*

| Member | Function | Center |
|---|---|---|
| Dr. L. Timmers | Voorzitter registratiecommissie Interventiecardioloog | St. Antonius Ziekenhuis |
| Dr. B.J.L. van den Branden | Interventiecardioloog | Amphia Ziekenhuis |
| Dr. R. Delewi | Interventiecardioloog | Amsterdam UMC |
| Prof. dr. W.A.L. Tonino | Interventiecardioloog | Catharina Ziekenhuis |
| Prof. dr. N.M.D.A. van Mieghem | Interventiecardioloog | Erasmus Medisch Centrum |
| Dhr. C.E. Schotborgh | Interventiecardioloog | HagaZiekenhuis |
| Dr. R.S. Hermanides | Cardioloog | Isala |
| Dhr. F. van der Kley | Interventiecardioloog | Leids Universitair Medisch Centrum |
| Dr. P. Vriesendorp | Cardioloog | Maastricht UMC+ |
| Dhr. F. Porta | Cardiothoracaal Chirurg | Medisch Centrum Leeuwarden |
| Dhr. K.G. van Houwelingen | Interventiecardioloog | Medisch Spectrum Twente |
| Dr. G. Amoroso | Interventiecardioloog | Onze Lieve Vrouwe Gasthuis |
| Dr. M. van Wely | Interventiecardioloog | Radboudumc |
| Prof. dr. M. Voskuil | Interventiecardioloog | UMC Utrecht |
| Dhr. H.W. van der Werf | Interventiecardioloog | Universitair Medisch Centrum Groningen |